\documentclass[12pt,preprint]{aastex}

\shorttitle{X-shaped Ribbon Flare}
\shortauthors{Li et al.}

\begin{document}

\title{Observations of an X-shaped Ribbon Flare in the Sun and Its Three-dimensional Magnetic Reconnection}

\author{Y. Li$^{1,2,3}$, J. Qiu$^{2}$, D. W. Longcope$^{2}$, M. D. Ding$^{1,3}$, K. Yang$^{1,3}$}
\affil{$^1$School of Astronomy and Space Science, Nanjing University, Nanjing 210093, China}
\affil{$^2$Department of Physics, Montana State University, Bozeman, MT 59717, USA}
\affil{$^3$Key Laboratory for Modern Astronomy and Astrophysics (Nanjing University), Ministry of Education, Nanjing 210093, China}
\email{yingli@nju.edu.cn}

\begin{abstract}
We report evolution of an atypical X-shaped flare ribbon which provides novel observational evidence of three-dimensional (3D) magnetic reconnection at a separator. The flare occurred on 2014 November 9.  High-resolution slit-jaw 1330 \AA\ images from the {\em Interface Region Imaging Spectrograph} reveal four chromospheric flare ribbons that converge and form an X-shape. Flare brightening in the upper chromosphere spreads along the ribbons toward the center of the ``X'' (the X-point), and then spreads outward in a direction more perpendicular to the ribbons. These four ribbons are located in a quadrupolar magnetic field. Reconstruction of magnetic topology in the active region suggests the presence of a separator connecting to the X-point outlined by the ribbons. The inward motion of flare ribbons in the early stage therefore indicates 3D magnetic reconnection between two sets of non-coplanar loops that approach laterally, and reconnection proceeds downward along a section of vertical current sheet. Coronal loops are also observed by the Atmospheric Imaging Assembly on board the {\em Solar Dynamics Observatory} confirming the reconnection morphology illustrated by ribbon evolution.

\end{abstract}


\section{Introduction}
\label{intro}

Magnetic reconnection refers to topological reconfiguration of magnetic field, which is widely accepted as the mechanism of energy release in solar flares \citep{prie02}. In the standard two-ribbon flare model \citep{carm64,stur68,hira74,kopp76}, oppositely-directed magnetic field lines reconnect at a current sheet, creating closed field lines that cross over a polarity inversion line (PIL) below the current sheet and a plasmoid structure above. More sets of anti-parallel field lines flow into the current sheet and reconnect at a growing height, producing two flare ribbons moving outward away from the PIL. In this classical two-dimensional (2D) picture, magnetic reconnection takes place at a null-line (or X-line) between anti-parallel coplanar field lines, forming a post-flare arcade and ribbons outlining the feet of the arcade.

In reality, magnetic reconnection in solar flares must be three-dimensional (3D). In the 3D framework, magnetic reconnection takes place preferentially at a separator (intersection of two separatrices; \citealt{baum80,lauy90,long96,long05}) or at its generalization, a hyperbolic flux tube (HFT; intersection of two quasi-separatrix layers, or QSLs; \citealt{demo96,demo97,tito02}). The 3D magnetic reconnection is manifest in many two-ribbon flare observations \citep{gorb88,mand91,demo93,demo07}, such as bi-directional spread of post-flare arcade \citep{aula07}, anti-parallel motions of hard X-ray sources \citep{aula06,demo07}, elongation motion of flare ribbons \citep{qiuj02,qiuj10,qiuj09,flet04}, and $J$-shaped flare ribbons \citep{aula12}. Besides two-ribbon flares, circular-ribbon flares also exhibit a special configuration of 3D magnetic reconnection with a null point and a fan-spine structure \citep{mass09,sunx13,yang15}.

In this Letter, we present observational evidence for a different scenario from those reported previously: an X-shaped ribbon flare with the ribbons intersecting at the center, revealed by the high-resolution ultraviolet images from the {\em Interface Region Imaging Spectrograph} ({\em IRIS}; \citealt{depo14}). Simultaneously, the extreme-ultraviolet images from the Atmospheric Imaging Assembly (AIA; \citealt{leme12}) on board the {\em Solar Dynamics Observatory} ({\em SDO}) show converging non-coplanar loops. These elements suggest a 3D reconnection scenario with a separator, and for the first time, that separator and the reconnection occurring there have been observed.

\section{Observations}

The M2.3 flare presented here was observed by {\em IRIS} in slit-jaw 1330 \AA~images (SJI 1330 \AA) from 15:17 UT to 16:05 UT on 2014 November 9. Figure \ref{fig-obs} gives an overview of the flare. The flare 1--8 \AA~soft X-ray emission starts at $\sim$15:22 UT and peaks at 15:32 UT, as shown in the top panel. Also shown are light curves of different parts of the flare ribbons in SJI 1330 \AA, which observe the upper chromosphere and lower transition region. 

The lower panels of the figure and the animation show the flare evolution in SJI 1330 \AA, and AIA 1600, 171, and 131 \AA~passbands, demonstrating formation of an X-shaped flare ribbon. Brightenings along four branches of the ribbon in the northeast (NE), southeast (SE), northwest (NW), and southwest (SW) spread toward each other and finally cross at one point around the flare peak, when the ribbons exhibit an X-shape. Furthermore, we see evident post-flare loops in AIA 131 and 171 \AA~images in the decay phase, which appear to converge toward the center of the X-shape.

Here, we analyze flare ribbon evolution observed in SJI 1330 \AA~which have a pixel size of $0.\!\!^{\prime\prime}166$ and a cadence of 37~s. We also use the AIA and HMI (Helioseismic and Magnetic Imager; \citealt{scho12}) data from {\em SDO} to construct magnetic topology of the flare region. The AIA images have a pixel scale of $0.\!\!^{\prime\prime}6$. The HMI magnetic field data used here are taken from pre-flare. All the images from {\em IRIS} and {\em SDO} are co-aligned by comparing the sunspot features shown in both SJI 2832 \AA~and AIA 1700 \AA~images. 

\section{Analysis and Results}

\subsection{Flare Ribbon Motion}
 
This flare shows an unusual ribbon pattern as revealed by high spatial-resolution SJI 1330 \AA. The ribbon's apparent motion reveals three stages of flare evolution: the initial (15:22--15:28 UT), main (15:28--15:32 UT), and last (15:32--15:38 UT) stages, denoted by the red vertical dotted lines in the top panel of Figure \ref{fig-obs}. The top panels of Figure \ref{fig-ribbon} present the three-stage motion pattern of footpoint brightenings in 1330 \AA. The SJI 1330 \AA~observe brightening in the flare ribbons as well as some low-temperature loops. To distinguish brightening of footpoints from loops or other transient non-ribbon features, we select flaring pixels whose intensity is enhanced to be more than 25 times the intensity of the quiescent-Sun for more than two minutes \citep{long07}, and identify these pixels as footpoints of flare loops. 

Seen from Figure \ref{fig-ribbon}, in the initial stage, two ribbons on the right of the flare region (NW ribbon in positive magnetic fields and SW ribbon in negative fields) brighten first and generally show an elongation motion toward the left. These ribbons are on the two sides of the PIL and are nearly parallel with the PIL along the east-west direction, and their morphology resembles the standard two-ribbon flare configuration. In the subsequent main stage, the other two ribbons on the left (NE ribbon in negative fields and SE ribbon in positive fields) are brightened from the outer-most ends, and the brightenings spread to the right as well as approach each other. In the meanwhile, the previously brightened NW and SW ribbons spread further, converging toward each other. The four ribbons, when fully formed, intersect at one point marked in the top panels of Figure \ref{fig-ribbon}, and form an unusual X-shape extending from this point, which we refer as the X-point in the foregoing text. Since the X-point is located at the center of a quadrupolar magnetic structure, it is plausible that a separator may be anchored there. In the last stage, all the ribbons display an outward motion; in particular, the NW and SW ribbons move outward away from the local PIL like in the 2D picture.

We track the ribbon brightening to measure the apparent motion, which is then shown in the two middle panels of Figure \ref{fig-ribbon}. For the inward motion occurring in the initial and main stages, we measure the distance of the ribbon fronts from the X-point. In the initial stage, the fronts of the NW and SW ribbons both move toward the X-point but in the direction nearly parallel to the PIL separating the two ribbons. At the end of this stage, the NW ribbon is $\sim$3 Mm from the X-point. In the following main stage, the fronts of the NW and SW ribbons have changed the motion direction to also approach each other until they converge at the X-point; the apparent speed of the converging motion is $\sim$17 and $\sim$85 km s$^{-1}$ for the NW and SW ribbons, respectively. Simultaneously, the NE and SE ribbons are brightened and their fronts converge to the X-point at the speeds of $\sim$47 and $\sim$66 km s$^{-1}$, respectively. The four ribbons intersect at the X-point at 15:31 UT. 

In the main stage, whereas the leftmost fronts of the NW and SW ribbons are moving toward the X-point, the previously-formed portions of these ribbons also start to spread outward away from the PIL. This separation motion continues into, and dominates, the last stage. To measure the separation motion of NW and SW ribbons, we track the distance of the ribbon fronts along the direction perpendicular to the PIL at a point marked by a triangular symbol in the top left panel of Figure \ref{fig-ribbon}. The speed of this separation motion is much smaller, of mostly $<$ 9 km s$^{-1}$.

Apparent motion of flare ribbons is a consequence of energy release along flare loops successively formed by magnetic reconnection. Therefore, magnetic reconnection flux can be measured by summing up magnetic flux swept up by flare ribbons \citep{forb84,pole86}. The bottom panel of Figure \ref{fig-ribbon} shows reconnection flux and its time derivative, the reconnection rate, measured in the positive and negative fields respectively. It is seen that the reconnection flux grows rapidly in the main stage, which is dominated by the ribbon's inward motion. The total reconnection flux approaches 10$^{21}$ Mx with the positive and negative fluxes reasonably balanced. The reconnection rate reaches its maximum of 8$\times$10$^{18}$ Mx s$^{-1}$ early in the main stage and diminishes toward the end of this stage. All of these suggest that most of reconnection and subsequent energy release take place during the main stage of the flare, when ribbons form the peculiar X-shape.

\subsection{Magnetic Topology by 3D Field Extrapolation}

According to a common topological picture\footnote{Here we refer only to the interpretation of this flare, in which we find evidence for separator reconnection. We recognize that some other events show evidence for reconnection associated with more general topological features like QSLs or HFTs.}, flare ribbons form at the bases of separatrices of the coronal field. A positive ribbon separates the positive polarity connecting to one negative source, say $N1$, from that connecting to another, $N2$. Similarly, the negative ribbon separates field lines originating at $P1$ from those originating at $P2$. The reconnection responsible for the flare occurs as field lines from two connectivities, $P1$--$N1$ and $P2$--$N2$, are eliminated to form field lines of the other two: $P1$--$N2$ and $P2$--$N1$. The increase of the latter two, and decrease in the former two, cause the prototypical spreading of the ribbons. The reconnection is supposed to occur at a separator formed by the intersection of the two separatrices. This is the one location where field lines from all four distinct connectivities are in close proximity. In the vast majority of cases, the separator\footnote{Note that numerous separators could be found in magnetic topological models \citep{long07a,parn10}.} is indirectly inferred, and believed to lie high in the corona, where the flare reconnection is actually occurring.

The present flare has the unusual property that its positive and negative flare ribbons cross one another at a point on the surface, namely the X-point. This suggests that the separatrices themselves intersect along a separator which extends all the way down to the solar surface. In this case, at least a portion of the reconnection is taking place at the lower boundary, where it is more readily observed.

To explore this hypothesis, we produced a topological model of the magnetic field using the Magnetic Charge Topology method \citep[MCT;][]{long05}. A section of the 14:34:13 UT line-of-sight HMI magnetogram, shown in the upper left panel of Figure~\ref{fig-model}, was extracted, and its flux partitioned into over 150 distinct polarity regions. These flux regions were replaced by magnetic point sources containing the same net flux and situated at their centroids. This approximation has been found to change the actual connectivity of the extrapolated field by no more than 15\%, but to render the field's topological features clear and easily identified \citep{long09}. From these point sources, we extrapolated a linear force-free field (LFFF) with $\alpha=+3\times10^{-10}\,{\rm cm}^{-1}$, in order to identify the connectivities, and the skeleton defining it. This value of $\alpha$ was chosen to produce field lines with the best overall resemblance to those observed in AIA 171 \AA~images.

The lower panels of Figure \ref{fig-model} show representative field lines of different connectivities from the LFFF extrapolation. Cyan lines show field lines from each of the pre-reconnection domains, while yellow lines show those field lines resulting from reconnection around the X-point. Magenta field lines show connections in other domains not immediately adjacent to the X-point. 

Red and blue curves in the lower panels show the surface manifestation of the separatrices whose intersection forms the separator. These same structures are over-plotted, with the same colors, in the upper panels as well. Green and violet curves in those panels show how these separatrices extend along the positive and negative polarities, respectively. The topological model of the separatrices can therefore be seen to follow the flare ribbons, conforming to our basic understanding of flare reconnection. In addition, the HMI magnetogram in Figure \ref{fig-model} shows that, near the X-point, the negative regions are clearly divided to be distinct sources, and the positive regions can be considered to be distinct sources as well, though the division is less prominent. The magnetic field with such a feature is inclined to the formation of a null point (and thus the X-point) and a genuine separatrix instead of a less distinct QSL. The separatrices cross at a photospheric null point located west and south of the observed X-point. Such a geometrical discrepancy is expected in a topological model of this kind, since a MCT model does not accurately model the actual field. We have then performed a LFFF extrapolation from the full-resolution, line-of-sight HMI magnetogram, rather than a point-source approximation of it (i.e.\ MCT), and found a null point just above the lower boundary, situated slightly closer to the X-point. A non-linear force-free field (NLFFF) extrapolation using pre-processed HMI vector data has shown a similar, low-lying null point, still closer to the actual location (X. Sun, private communications). In all cases the skeleton from the null point outlines the same topological structure. We expect the same will be true of the actual magnetic field, of which the MCT, LFFF and NLFFF, are decreasingly crude models.

The actual field, like the LFFF extrapolation, has distinct-connectivity families separated by separatrices whose intersection with the surface appears along the flare ribbons. In the present field, these separatrices intersect at a separator that extends down to the surface, therefore allowing reconnection to occur very near the solar surface. According to this picture, field lines are forged through the process of reconnection --- the post-reconnection flux --- and should be energized in the process. These field lines, shown in yellow and magenta in the lower panels of Figure \ref{fig-model}, match those coronal loops most evident in the hot AIA channels (e.g., 94 \AA) shown in the lower right panel of that figure. These images show little or no evidence of the other family of field lines (cyan lines), presumably because they are not energized by the flare.

In the LFFF model employed here, the separator connecting to the X-point, together with another separator also lying within the fan of the X-point-related null, are found and shown as cyan curves in the top panels of Figure 3. Each of the separators is a single field line along which the separatrix surfaces intersect transversally. In the actual field, pre-reconnection stressing will deform that line into a current-carrying ribbon \citep{long01,long02,long04}. This deformation resembles that observed in 2D models of X-points, which stretch into current sheets under stress \citep{syro71}. In either case, the structure elongates to create a surface across which pre-reconnection domains meet to reconnect. Reconnected flux is then expelled through the sheet, emerging at its tips where post-reconnection flux lies. Following this logic we expect the surface intersection of our separator to be stretched horizontally (i.e.\ east-west). This is, in fact, what SJI 1330 \AA~show during the most intense periods of reconnection.

\section{Conclusion \& Discussions}
\label{discussion}

In this Letter, we report an atypical X-shaped flare ribbon, its evolution, and magnetic topology. These observations present the first evidence, as far as we know, for a special scenario of separator reconnection. Reconstruction of magnetic topology in the active region suggests the presence of separators connecting to the X-point outlined by the ribbons. The inward motion of four flare ribbons as well as converging post-flare loops illustrate sequential 3D reconnection between two sets of non-coplanar loops that approach laterally. It is also evident that initiation of this sequence of 3D reconnection at the X-point is related to the flare development nearby.

\begin{figure*}[htb]
\begin{center}
\includegraphics[width=12cm]{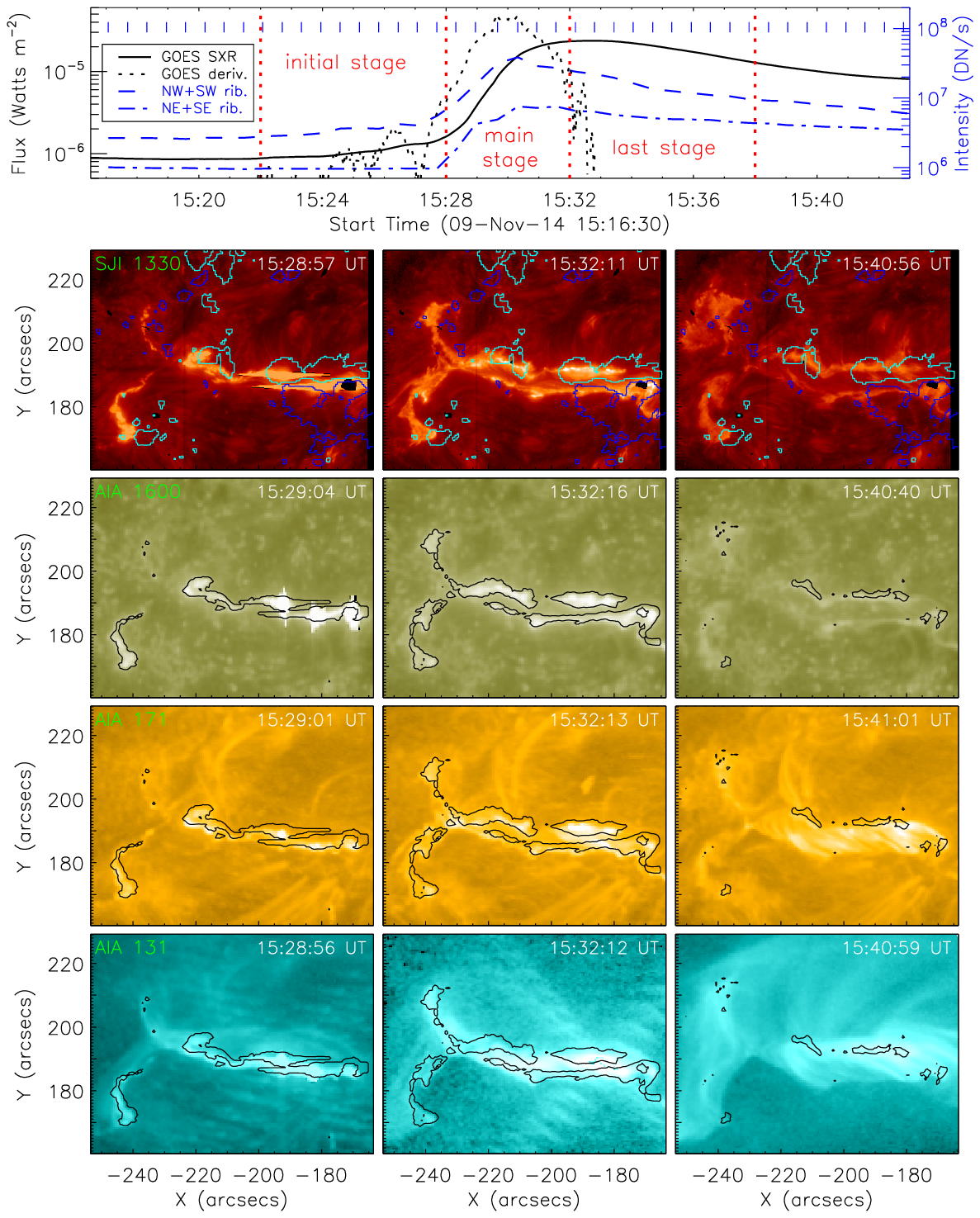}
\caption{Top panel: {\em GOES} 1--8 \AA~soft X-ray flux (left coordinate) and its derivative as well as SJI 1330 \AA~light curves (right coordinate) of the right (NW+SW) and left (NE+SE) parts of the ribbons for the M2.3 flare. The short vertical lines in blue mark the observing times of SJI 1330 \AA. The flare displays three evolution stages shown by the red vertical dotted lines. In the SJI 1330 \AA\ images, the contours represent the positive (cyan) and negative (blue) magnetic field at 500 and -500 G, respectively. The contours in the AIA 1600, 171, and 131 \AA~images show the SJI 1330 \AA~intensity of 25 times the average quiescent-sun intensity. (An animation of this figure is available.)} 
\label{fig-obs}
\end{center}
\end{figure*}

\begin{figure*}[htb]
\begin{center}
\includegraphics[width=11cm]{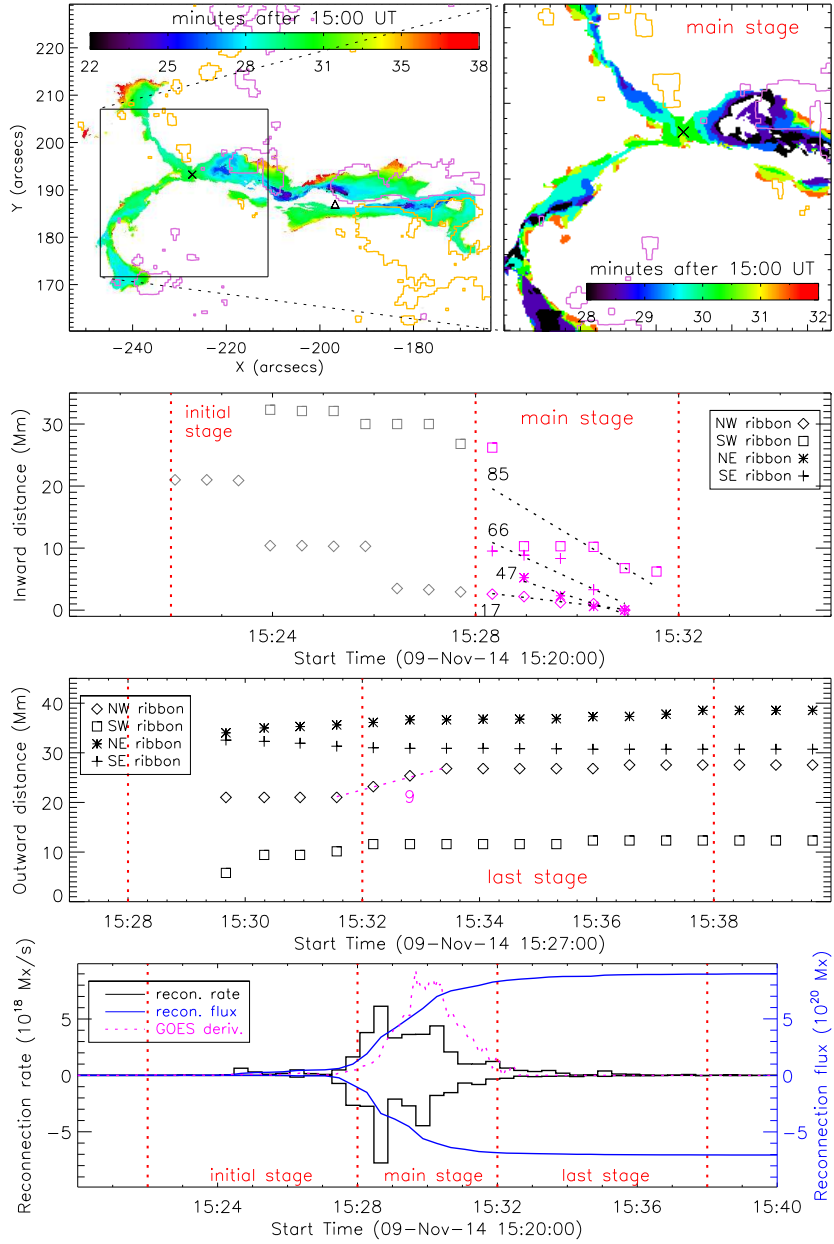}
\caption{Top row: evolution of the footpoint brightenings observed in SJI 1330 \AA\ in the full view (left) and zoomed-in view (right; black box in the left). Contours mark the positive (pink) and negative (orange) magnetic field at 500 and -500 G, respectively. Middle rows: distance of ribbon brightenings from the X-point (indicated by an X in the top row) in the initial and main stages, and then from a fixed point at the PIL (indicated by a triangle in the top left panel) in the last stage (see text). The numbers give the mean speeds (in units of km s$^{-1}$) of the apparent ribbon motion. The data points with diamond and square in the third row are multiplied by 6 for a better display. Bottom row: reconnection flux (blue) and reconnection rate (black) measured in positive and negative magnetic fields in the whole flaring region.}
\label{fig-ribbon}
\end{center}
\end{figure*}

\begin{figure*}[htb]
\begin{center}
\includegraphics[width=13cm]{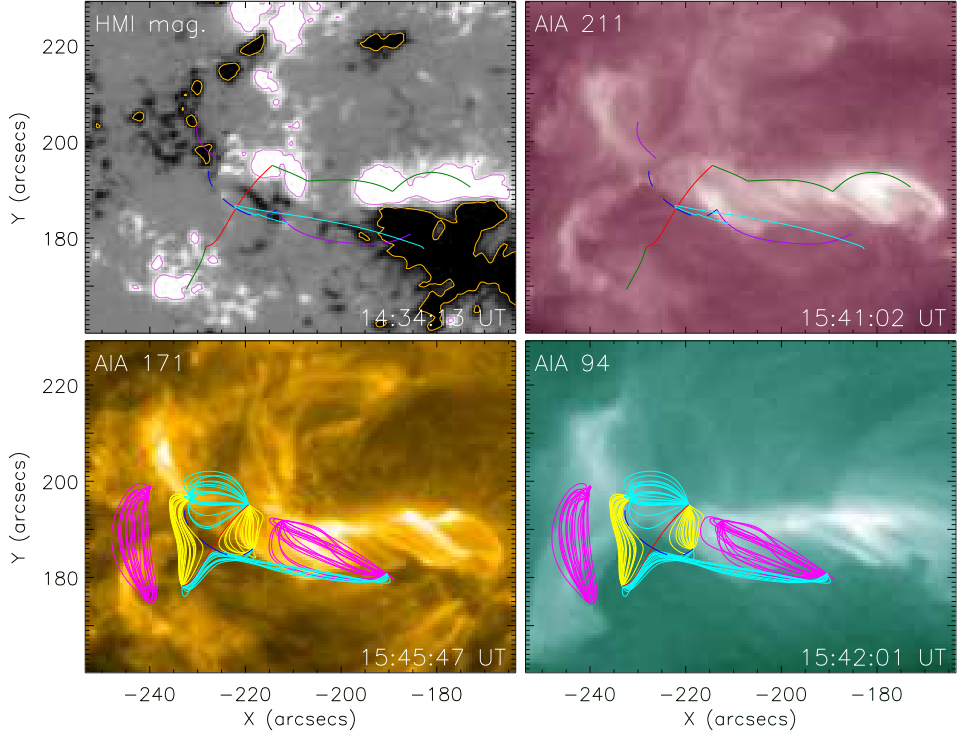}
\caption{The topological magnetic model superimposed on imaging data. The upper left panel shows a line-of-sight HMI magnetogram with contours representing the positive (pink) and negative (orange) polarities at 500 and -500 G respectively. The violet and blue lines show the traces of the negative separatrix, and green and red curves are the traces of the positive separatrix. The cyan lines show the separators lying within the fan surface of the null point and connecting to null points within the negative ribbon. The upper right panel shows an AIA 211 \AA~image along with the magnetic skeleton. The lower panels show some of the separatrix traces (red and blue) as well as representative field lines from several connections (cyan, yellow, and magenta), over AIA 171 and 94 \AA~images.}
\label{fig-model}
\end{center}
\end{figure*}

\begin{figure*}[htb]
\begin{center}
\includegraphics[width=15cm]{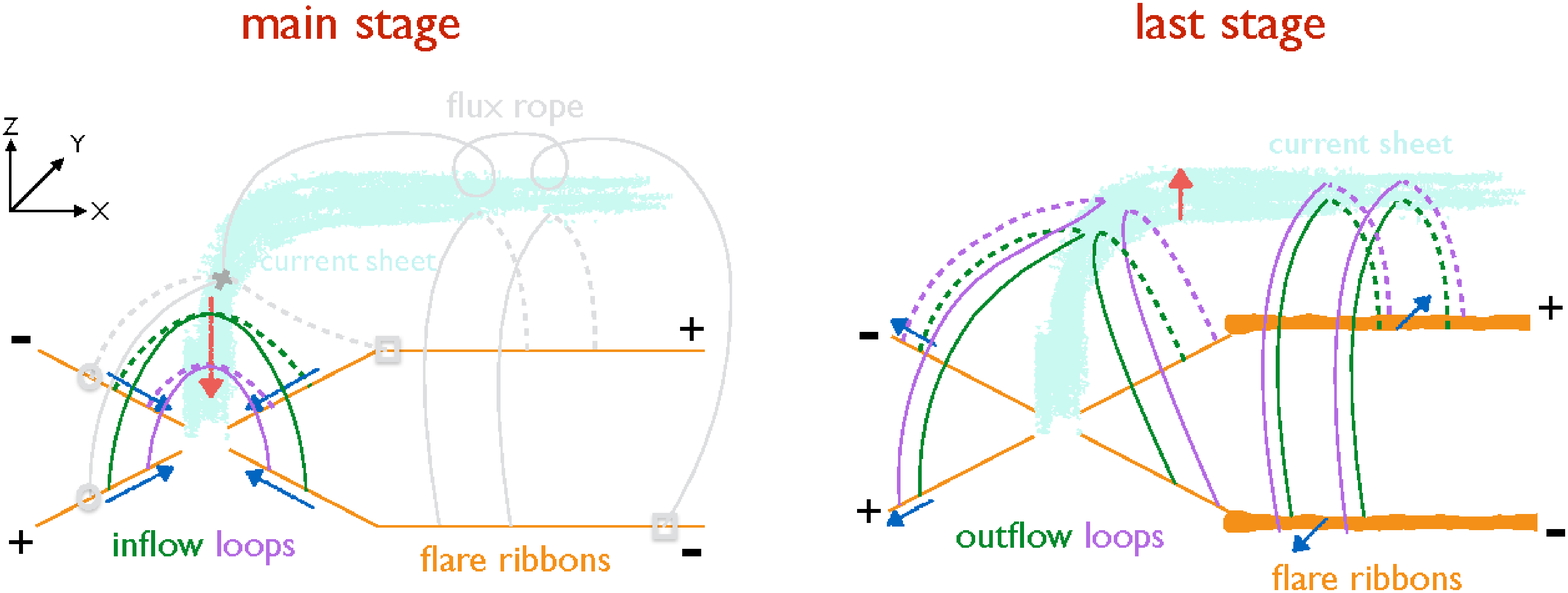}
\caption{Sketch of the 3D separator reconnection. The main stage is dominated by an inward ribbon motion (blue arrows) to the X-point due to the reconnection site moving downward (red arrow), which is perhaps triggered by a perturbed flux rope. Two sets of non-coplanar inflowing loops approach laterally and reconnect at the current sheet. In the last stage, the outward ribbon motion dominates with the reconnection site moving upward, generating outflowing (post-flare) loops.}
\label{fig-cartoon}
\end{center}
\end{figure*}

The reconnection configuration and evolution in this flare can be sketched in Figure \ref{fig-cartoon}, demonstrating a current sheet that extends from the right of the X-shaped ribbon first horizontally and then vertically into the lower atmosphere at the X-point. Prior to the formation of the X-shaped ribbon, reconnection starts and spreads leftward along the horizontal current, perhaps triggered by a disturbed overlying flux rope\footnote{The flux rope structure is visible in AIA 131 \AA~images (see the animation of Figure \ref{fig-obs}).}, forming the NW and SW ribbons nearly parallel to each other. Reconnection then proceeds downward along the vertical current with two sets of non-coplanar (inflow) loops reconnecting laterally (left panel). The unusual inward motion of the ribbons map the footpoints of outflow loops formed at progressively lower heights. As the flux rope expands (right panel), the reconnection site moves upward, now almost in a direction perpendicular to the horizontal current, producing progressively high-lying post-flare (outflow) loops as well as the apparent outward motion of the ribbons (similar to the standard 2D reconnection scenario). Such a two-stage reconnection evolution is reported for the first time for an X-ribbon flare occurring along a curved separator line in a complex magnetic field.


\acknowledgments
The authors thank the anonymous referee for constructive comments to improve the manuscript.  We thank Xudong Sun for helpful discussions including results of his NLFFF modeling. This project was supported by NSFC under grants 10933003, 11373023, and 11403011, and by NKBRSF under grants 2011CB811402 and 2014CB744203. Y.L. is also supported by the Postdoctoral Science Foundations from Jiangsu Province and China Postdoctoral Office, and by the Fundamental Research Funds for the Central Universities. The work at MSU is also supported by the NSF grant 1460059. {\em IRIS} is a NASA small explorer mission. \textit{SDO} is a mission of NASA's Living With a Star Program.

\bibliographystyle{apj}

\end{document}